\documentclass[a4paper,10pt,twoside]{cpc-hepnp}
\usepackage{multicol}
\usepackage{graphicx}
\usepackage{booktabs}
\usepackage{amssymb,bm,mathrsfs,bbm,amscd}
\usepackage[tbtags]{amsmath}
\usepackage{lastpage}
\usepackage{subfigure}

\begin{document}

\fancyhead[c]{\small Submitted to 'Chinese Physics C'}
\fancyfoot[C]{\small 010201-\thepage}

\footnotetext[0]{Received 31 June 2015}

\title{ Injection method of barrier bucket supported by off-aligned electron cooling for CRing of HIAF\thanks{Supported by the new interdisciplinary and advanced pilot fund of Chinese Academy of Sciences }}

\author{%
      Guo-Dong Shen (Éê¹ú¶°)$^{1,2;1)}$\email{shenguodong@impcas.ac.cn}%
\quad Jian-Cheng Yang (Ñ³É)$^{1}$
\quad Jia-Wen Xia (ÏļÑÎÄ)$^{1}$\\
\quad Li-Jun Mao (ðÁ¢¾ü)$^{1}$
\quad Da-Yu Yin (Òó´ïîÚ)$^{1}$
\quad Wei-Ping Chai (²ñΰƽ)$^{1}$
\quad Jian Shi (ʯ½¡)$^{1}$\\
\quad Li-Na Sheng (Ê¢ÀöÄÈ)$^{1}$
\quad A. Smirnov$^{3}$
\quad Bo Wu (ÎⲨ)$^{1,2}$
\quad He Zhao (ÕÔºØ)$^{1,2}$
}
\maketitle

\address{%
$^1$ Institute of Modern Physics, Chinese Academy of Sciences, Lanzhou 730000, China\\
$^2$ University of Chinese Academy of Sciences, Beijing 100049, China\\
$^3$ Joint Institute for Nuclear Research, Dubna 141980, Russian Federation\\
}

\begin{abstract}
A new accelerator complex, HIAF (the High Intensity Heavy Ion Accelerator Facility), has been approved in China. It is designed to provide intense primary and radioactive ion beams for research in high energy density physics, nuclear physics, atomic physics as well as other applications. In order to achieve a high intensity of up to $5\times10^{11}$ ppp $^{238}$U$^{34+}$, the Compression Ring (CRing) needs to stack more than 5 bunches transferred from the Booster Ring (BRing). However, the normal bucket to bucket injection scheme can only achieve an intensity gain of 2, so an injection method, fixed barrier bucket (BB) supported by electron cooling, is proposed. To suppress the severe space charge effect during the stacking process, off-alignment is adopted in the cooler to control the transverse emittance. In this paper, simulation and optimization with the BETACOOL program are presented.
\end{abstract}



\footnotetext[0]{\hspace*{-3mm}\raisebox{0.3ex}{$\scriptstyle\copyright$}2013
Chinese Physical Society and the Institute of High Energy Physics
of the Chinese Academy of Sciences and the Institute
of Modern Physics of the Chinese Academy of Sciences and IOP Publishing Ltd}%

\begin{multicols}{2}

\section{Introduction}

HIAF (the High Intensity Heavy Ion Accelerator Facility) is a new accelerator facility under design at the Institute of Modern Physics (IMP), Chinese Academy of Sciences, to provide high intensity heavy ion beams for a wide range of experiments in high energy density physics, nuclear physics, atomic physics and other applications~\cite{ref1}. It consists of two Superconducting Electron-Cyclotron-Resonance ion sources (SECR) and a high intensity H$^2_+$ ion source (LIPS), a 17 MeV/u superconducting Ion Linac (iLinac), a 34 Tm Booster Ring (BRing), a 43 Tm multifunctional Compression Ring (CRing), a high precision Spectrometer Ring (SRing), an electron machine based on an Energy Recovery Linac (ERL), and several experiment terminals. Its schematic layout is shown in Fig.~\ref{fig1}.

Generally the beam injection between synchrotrons is performed by a bucket to bucket scheme. The circumferences of the two synchrotrons have an integer ratio and the RF systems operate at the same frequency. Thus injected bunches match buckets well and the filamentation phenomenon in longitudinal phase space can be suppressed effectively~\cite{ref2}. However, this scheme has the restriction that the intensity gain cannot exceed the circumference ratio. The injection method, fixed barrier bucket supported by electron cooling, can exploit the longitudinal phase space significantly. A barrier bucket (BB) RF system builds spaces along the beam direction time and time again, enabling injection of many more bunches~\cite{ref3}. Electron cooling is essential to shrink the momentum deviations and make the barrier bucket stacking more efficient~\cite{ref4}. In the stacking process, overcooling of transverse emittance can lead to a severe space charge effect. Particles will suffer instability and
 \begin{center}
\includegraphics[width=1.0\columnwidth]{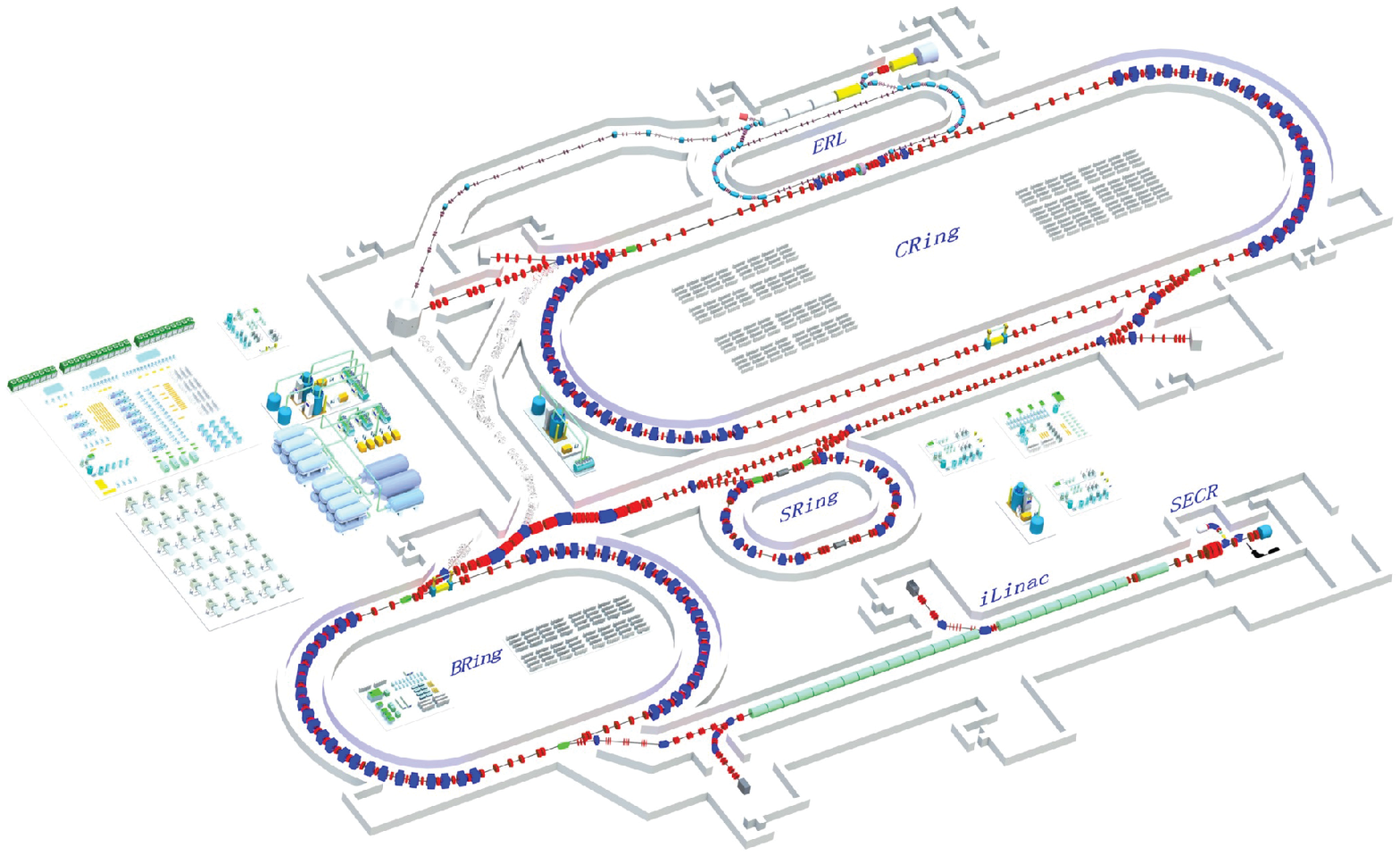}
\figcaption{\label{fig1} Overview of the HIAF facility. }
\end{center}
 get lost when their tunes are pushed near resonance lines. So a small angle between electron beam and ion orbit, namely off-alignment, is introduced to control overcooling~\cite{ref5}. In this paper, the conceptual design of the stacking scheme and the simulation result are reported.

\section{Stacking scheme}

The fixed barrier bucket RF system generates two isolated quasi-sinusoidal shape pulses with opposite voltages in one revolution. A local potential well is formed in longitudinal phase space. After sufficient cooling particles stay in the bottom of the well and leave the flat-top empty. Thus the next bunch can be injected here~\cite{ref6,ref7,ref8}. The depth of the potential well, also known as the barrier height, is defined as~\cite{ref9}
\begin{eqnarray}
\label{eq1}
\left( \frac{dp}{p} \right)_{height} = \frac{1}{\beta}\sqrt{\frac{ZeV_{peak}}{\eta\gamma Am_0c^2}\frac{\Delta\varphi}{\pi}}.
\end{eqnarray}
where $Z$ is the charge state, $V_{peak}$ is the peak barrier voltage, $\beta$ and $\gamma$ are relative parameters, $m_0$ is the average nucleon mass in the rest frame, $c$ is the velocity of light, $\eta$  is the slip factor and $\Delta\varphi$ is the barrier phase width. The longitudinal phase space is divided into two parts, the stable area and unstable area, which correspond to the well and flat-top respectively. The stable area is used for storage of particles with small momentum deviations while the unstable area is for injection.

The schematic procedure of CRing stacking (as shown in Fig.~\ref{fig2}) can be described in 3 steps:
\begin{enumerate}
\item The bunch extracted from BRing propagates through the transfer line HFRS and gets to the unstable area. In the next several thousand turns, particles spread immediately to all over the longitudinal dimension. In the stable area momentum deviations become larger and particles shift faster due to the transfer of potential to kinetic energy.
\item Momentum deviations decrease due to the interaction between ion beam and electron beam. Particles fall into the  potential well one after another. The shrink of the transverse phase space shown in Fig.~\ref{fig2}(b) indicates that the transverse Courant-Snyder (CS) invariants move towards a nonzero value. This special phenomenon results from the nonzero balanced emittance of the off-aligned cooling effect, which will be explained in Section 3 at length.
\item	As shown in Fig.~\ref{fig2}(c) most of the particles are restricted in the stable area after sufficient cooling. The transverse CS invariants are concentrated in a narrow area. At this time the empty unstable area is ready to accept the next injection.
\end{enumerate}

\begin{center}
\includegraphics[width=1.0\columnwidth]{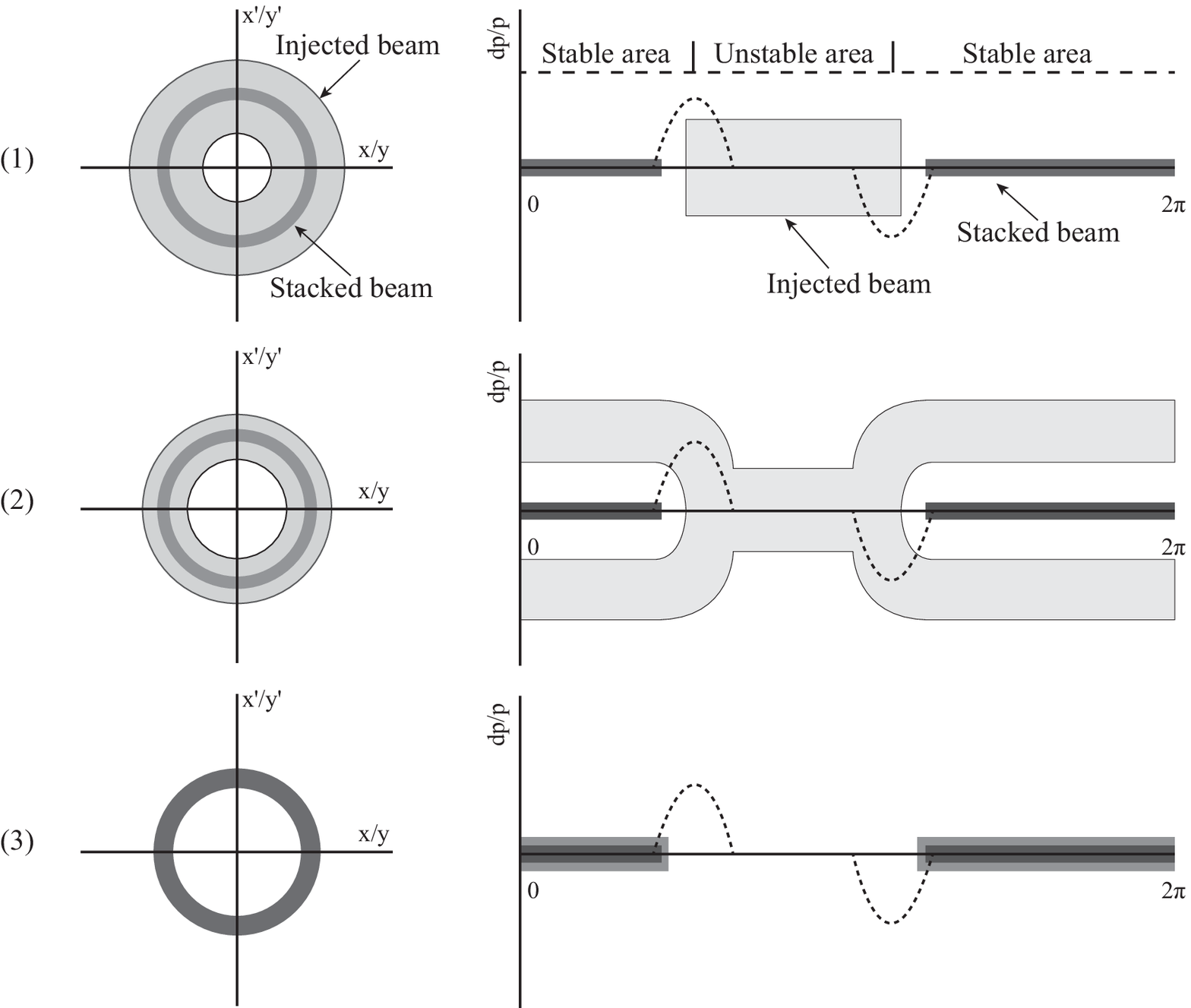}
\figcaption{\label{fig2}   Evolution of transverse phase space (left) and longitudinal (right) phase space during CRing Stacking. }
\end{center}

\section{Off-aligned electron cooling}
\label{sec3}

Electron cooling is critical to cool the longitudinal phase space down. However, it acts on the transverse dimension simultaneously due to its intrinsic characteristics~\cite{ref10}. The transverse emittance can be overcooled before being captured by BB, which gives rise to a grave space charge problem~\cite{ref11}.

Different from normal cooling, off-aligned electron cooling sets a small crossing angle between the electron beam and ion orbit. Particles get an extra heating friction effect. The final emittance is a balance between heating and cooling, which is described as a Hopf bifurcation phenomenon~\cite{ref12}. To lower the space charge effectively, off-alignment is set equal in the horizontal and vertical directions. Fig. \ref{fig3} shows the relationship between crossing angle and final emittance. The emittance is seen to grow rapidly below 0.4 mrad because of bifurcation and grow slowly above 0.6 mrad due to the increasing relative velocity between ions and electrons.

\begin{center}
\includegraphics[width=0.8\columnwidth]{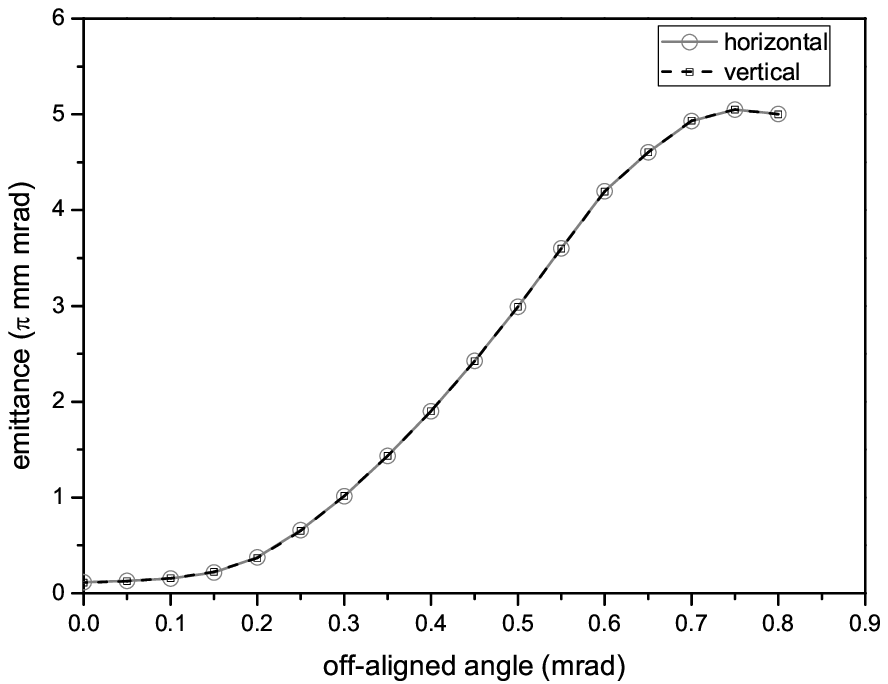}
\figcaption{\label{fig3}   The balanced transverse emittance under different off-alignments. }
\end{center}

\section{Stacking study and optimization}

The simulation is performed by the BETACOOL program, which takes account of several effects, including barrier bucket dynamics, off-aligned electron cooling, the IBS effect, recombination in the cooler, acceptance and injection kicker~\cite{ref13,ref14}. It should be noted that the space charge effect is not included in the simulation process, as the stacking process (36 seconds) is too long and unrealistic for 3D space charge effect tracking. The estimation of influence of space charge is performed in another way. The Laslett tune shift is calculated according to the particle distribution at the end of the simulation. If it is larger than the threshold value the influence of space charge is acceptable.

The 15 m electron cooler can provide electron beams with a wide energy range from 100 keV to 800 keV depending on the ion beam used. The electron gun can provide an electron current of 3 A with a round cross-section and uniform distribution. The main parameters are listed in Table~\ref{tab1}. The cooling force calculation is based on the Parkhomchuk semi-empirical formula, which is in quite reasonable agreement with available experimental results~\cite{ref15,ref16}:
\begin{eqnarray}
\label{eq2}
\vec{F} = -4\pi Z^2n_er^2_em_ec^2L_P\frac{\vec{V}}{(V^2+\Delta^2_{e,eff})^{3/2}}.
\end{eqnarray}
where all the variables are in the particle reference frame (PRF), $n_e$ is the density of the electron beam, $r_e$ is the classical electron radius, $c$ is the speed of light, $L_P$ is the Coulomb logarithm related to electron beam temperature, $V$ is the ion velocity, and $\Delta_{e,eff}$ is the effective electron velocity.

\begin{center}
\tabcaption{ \label{tab1}  Main parameters of the cooler.}
\footnotesize
\begin{tabular*}{80mm}{c@{\extracolsep{\fill}}ccc}
\toprule Parameter & Value \\
\hline
Cooling section length & 15m \\
Longitudinal magnetic field & 0.2T \\
$\beta_x/\beta_y$ & 20m/20m \\
$\alpha_x/\alpha_y$ & 0/0 \\
Transverse temperature & 0.5eV \\
Longitudinal temperature & 1meV \\
Effective temperature & 1meV \\
Electron beam radius & 1.8cm \\
Electron beam current & ${<}$3A \\
Neutralisation & 20\% \\
\bottomrule
\end{tabular*}
\vspace{0mm}
\end{center}
\vspace{0mm}

The accumulation scheme of BRing is under optimization. In the simulation, the parameters of the injected beam listed in Table~\ref{tab3} are quite conservative.

\begin{center}
\tabcaption{ \label{tab3}  Parameters of the injected beam.}
\footnotesize
\begin{tabular*}{80mm}{c@{\extracolsep{\fill}}ccc}
\toprule Parameter & Value \\
\hline
Horizontal emittance $\varepsilon_{H,rms}$ & 2.1$\pi$ mm mrad \\
Vertical emittance $\varepsilon_{V,rms}$ & 2.1$\pi$ mm mrad \\
Bunch length & 160m \\
Momentum spread& 5e-4\\
\bottomrule
\end{tabular*}
\vspace{0mm}
\end{center}
As the cycle of BRing is 6 s, 6 injections are performed in 36 s to achieve the stacking goal. The following parts will focus on the investigations of relevant parameters.

\subsection{Cooling current}

Electron cooling improves the beam quality through Coulomb collision between ions and electrons, so higher electron current brings a higher collision probability and a faster cooling. However the beam loss due to the recombination effect also increases. Fig.~\ref{fig4} shows the stacking efficiency under different electron beam currents. The cooling effect saturates when the current exceeds 1 A. Then the stacking efficiency increases slowly and even begins to drop due to the recombination effect.

\begin{center}
\includegraphics[width=0.8\columnwidth]{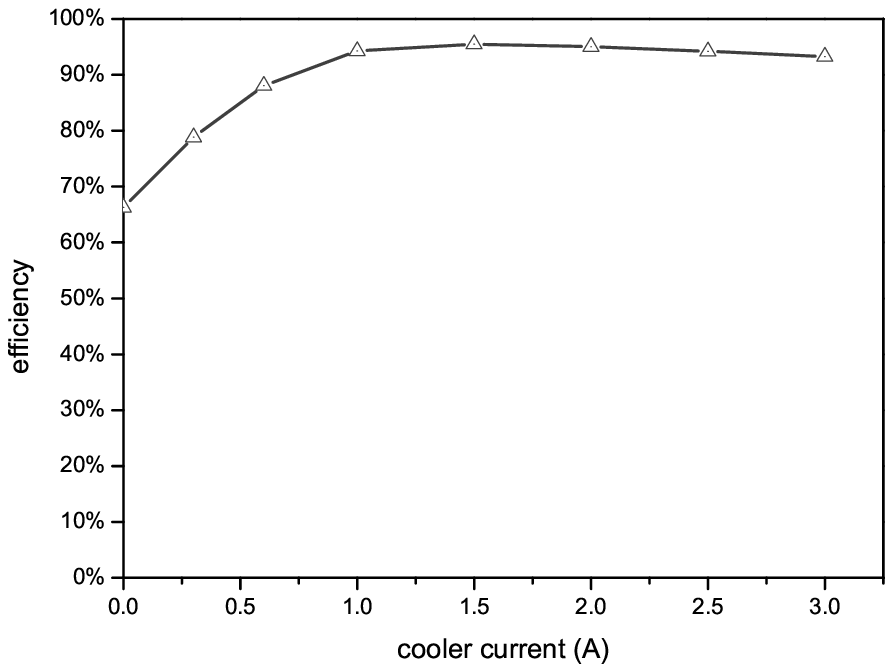}
\figcaption{\label{fig4}   The relationship between stacking efficiency and cooling current. }
\end{center}

\subsection{Off-aligned angle}

Fig.~\ref{fig5} shows how stacking efficiency varies with the off-aligned angle increasing. The off-alignment increases the transverse emittance and lowers the longitudinal heating due to the IBS effect. Meanwhile, following the increase of emittance the cooling force decreases owing to the transverse velocity deviation. As a balance between electron cooling and the IBS effect\cite{ref17,ref18,ref19}, 0.55 mrad is the best choice.

\begin{center}
\includegraphics[width=0.8\columnwidth]{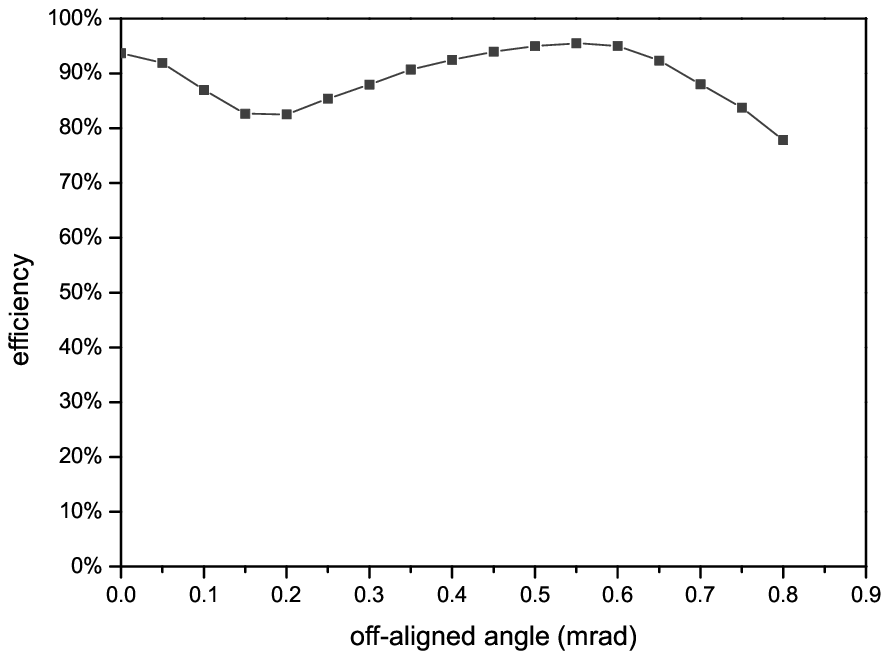}
\figcaption{\label{fig5}   The relationship between stacking efficiency and off-aligned angle. }
\end{center}

\subsection{BB voltage}

The cooled and stored particles are restricted in the potential well in longitudinal phase space. The deeper the potential well the better, because fewer particles can escape and be killed by the injection kicker. However, the injected particles get additional momentum deviations corresponding to the depth of the potential well in the stable area. This problem adds an extra burden to electron cooling. Fig.~\ref{fig6} indicates that 1000 V corresponding to a potential well of $2.47\times10^{-4}$ (the maximum relative momentum deviation can be restricted) is enough.

\begin{center}
\includegraphics[width=0.8\columnwidth]{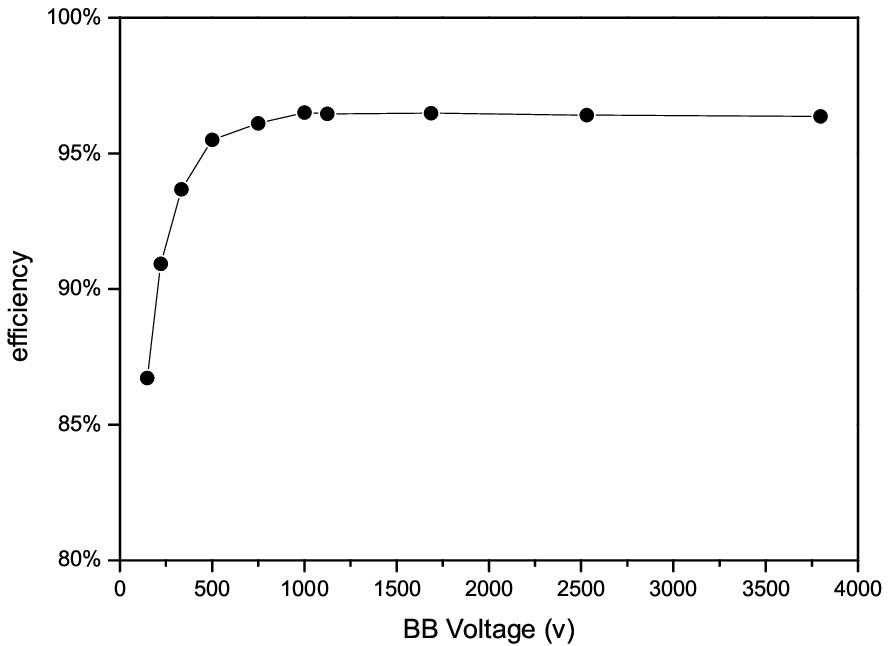}
\figcaption{\label{fig6}   The relationship between stacking efficiency and BB voltage. }
\end{center}

\subsection{Laslett tune shift}

Laslett tune shift is the average tune shift of particles due to the space charge effect. According to experience, particles may cross the resonance line and get lost when the Laslett tune shift is smaller than -0.1~\cite{ref20}. Off-alignment can increase the transverse emittance to lower the space charge effect, as shown in Fig.~\ref{fig7}. To keep the beam safe, the off-aligned angle should be larger than 0.5 mrad.

\begin{center}
\includegraphics[width=0.8\columnwidth]{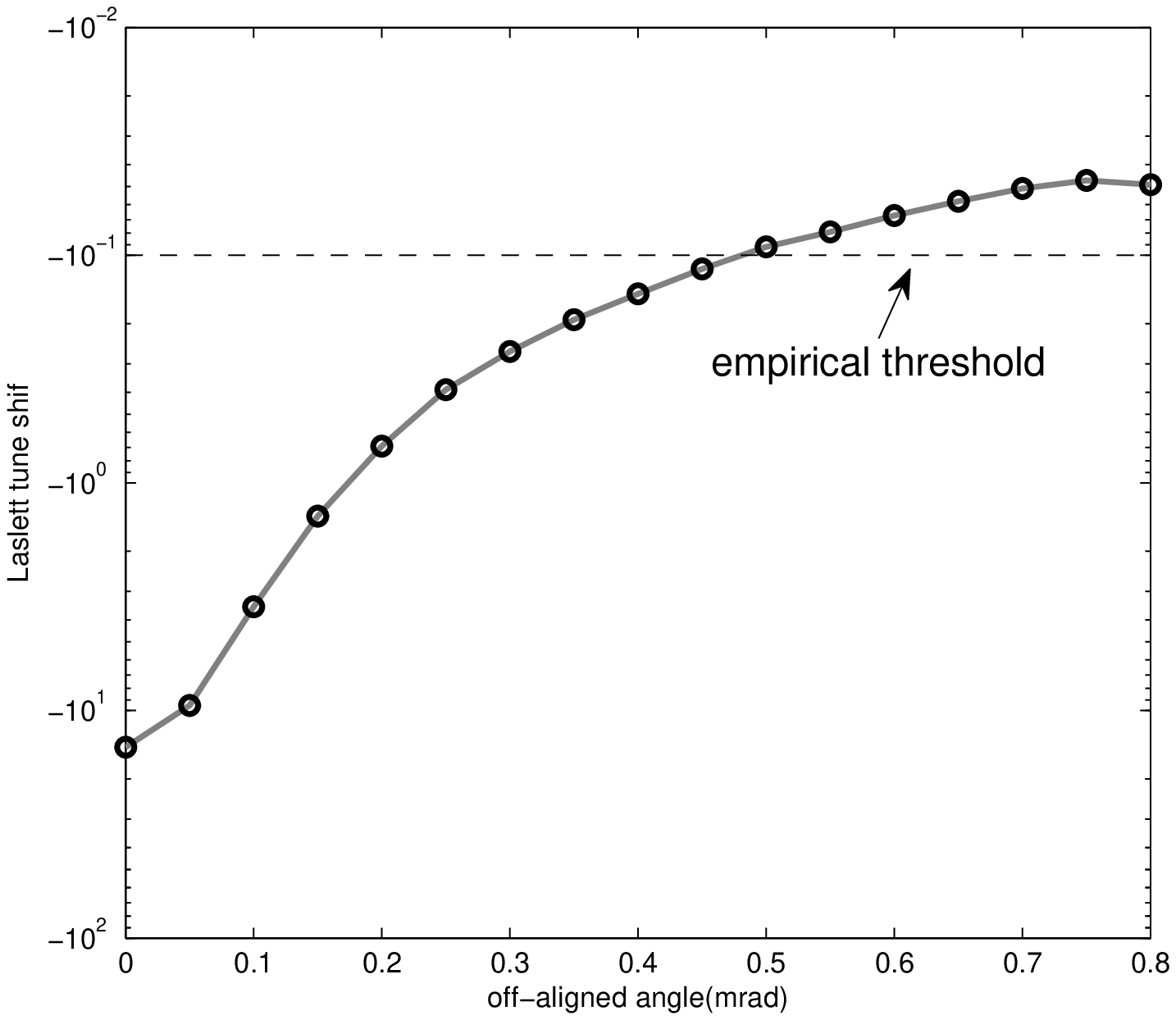}
\figcaption{\label{fig7}   The relationship between Laslett tune shift and off-aligned angle. }
\end{center}

\subsection{Optimization}

The parameters are optimized with the aim of maximizing the number of retained particles. The optimized parameters are listed in Table~\ref{tab2}. All of the results below are based on these. The optics parameters at the reference point are $\beta_x=\beta_y=10m$, $\alpha_x=\alpha_y=0$. Fig.~\ref{fig8} shows the particle distribution in horizontal phase space. All of the particles are concentrated on a narrow ellipse. This special phenomenon is due to the nonzero balanced emittance of the off-aligned cooing effect. Fig.~\ref{fig9} shows the beam distribution in transverse real space. Fig.~\ref{fig10} shows the distribution of momentum spread.

\begin{center}
\tabcaption{ \label{tab2}  The optimized parameters.}
\footnotesize
\begin{tabular*}{80mm}{c@{\extracolsep{\fill}}ccc}
\toprule Parameter & Value \\
\hline
Off-aligned angle & 0.55mrad \\
Cooling current & 1.5A \\
BB voltage & 1000V \\
retained particles & $5.79\times10^{11}$ \\
stacking efficiency & 96.5\% \\
Laslett tune shift & -0.078 \\
\bottomrule
\end{tabular*}
\vspace{0mm}
\end{center}
\vspace{0mm}

\begin{center}
\includegraphics[width=0.8\columnwidth]{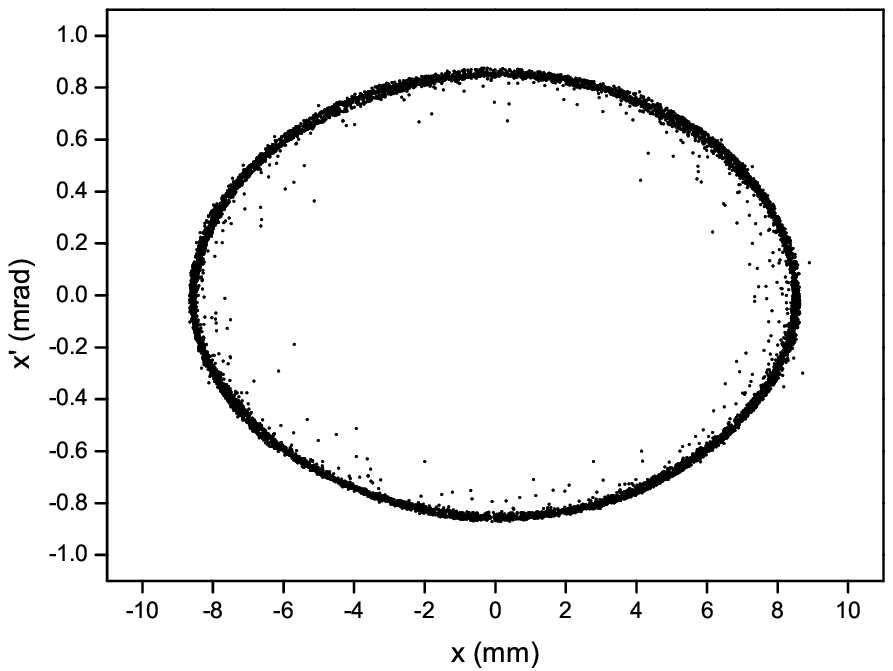}
\figcaption{\label{fig8}   Particle distribution in horizontal phase space. }
\end{center}

\begin{center}
\includegraphics[width=0.8\columnwidth]{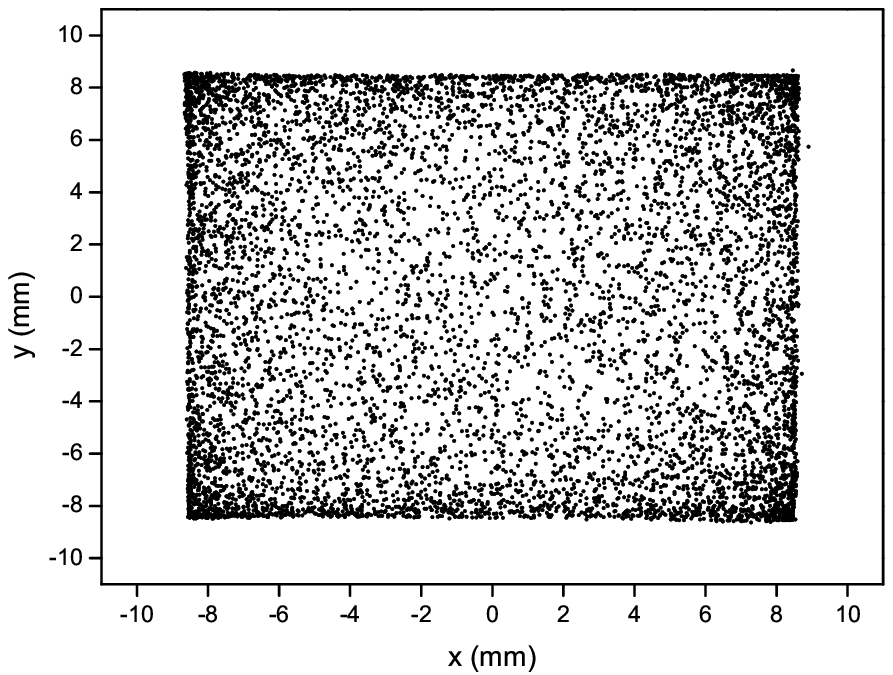}
\figcaption{\label{fig9}   Particle distribution in transverse real space. }
\end{center}

\begin{center}
\includegraphics[width=0.8\columnwidth]{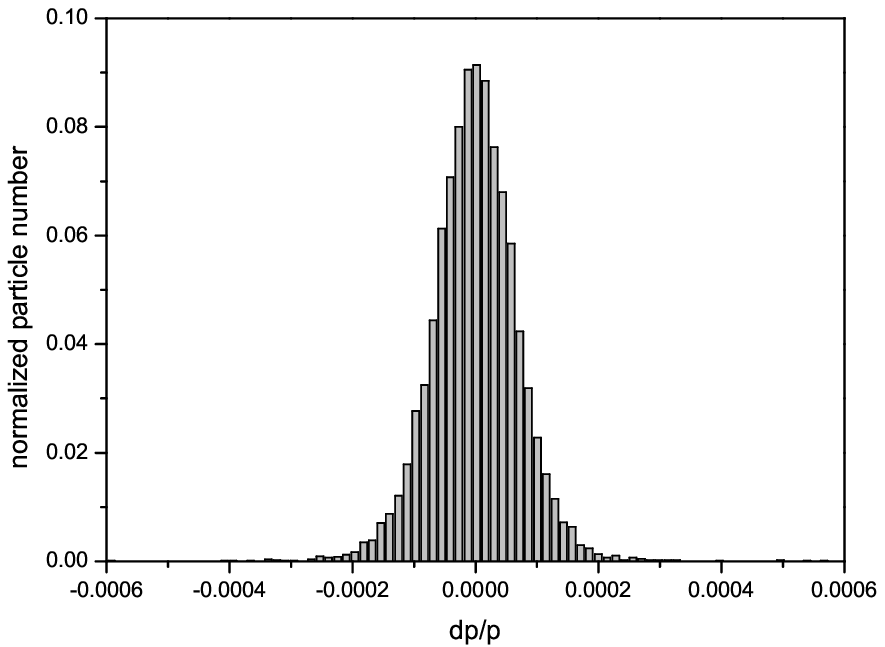}
\figcaption{\label{fig10}   Distribution of momentum deviations. }
\end{center}

\section{Conclusion}

 The simulation and optimization of BB stacking in the CRing at HIAF have been performed. The relations between stacking parameters and efficiency were researched based on the simulation code BETACOOL. The results demonstrate that BB stacking is an effective way to achieve intensity gain. With off-aligned electron cooling, the space charge effect is restricted remarkably in the high intensity case. The number of particles retained  after 6 injections reaches $5.79\times10^{11}$ ppp, corresponding to a stacking efficiency of 96.5\%. In future, the space charge effect in the stacking process will be further investigated.

\section{Acknowledgments}

The authors are very grateful for the valuable advice from A. Sidorin (JINR), and help from T. Katayama (Nihon University).

\end{multicols}

\clearpage

\vspace{15mm}


\begin{thebibliography}{90}

\vspace{3mm}

\bibitem{ref1}J. C. Yang, J. W. Xia, G. Q. Xiao et al, Nuclear Instruments and Methods in Physics Research Section B, 317: 263 (2013)

\bibitem{ref2}H. H. Gutbrod,  I. Augustin, H. Eickhoff et al, \emph{FAIR Baseline Technical Report} (Darmstadt: GSI, 2006), p. 429

\bibitem{ref3}J. E. Griffin, C. Ankenbrandt, J. A. MacLachlan et al, IEEE Trans. Nucl. Sci., NS-30(4): 3502 (1983)

\bibitem{ref4}T. Katayama, C. Dimopoulou, F. Nolden et al, Simulation study of barrier bucket accumulation with stochastic cooling at GSI ESR, in \emph{Proceedings of COOL'11}, edited by I. Meshkov (Alushta: JINR, 2011), p. 136

\bibitem{ref5}http://betacool.jinr.ru/reports/Uppsala\%20Uni\_2007\_March.pdf, retrieved 22th March 2016

\bibitem{ref6}M. Steck, C. Dimopoulou, B. Franzke et al, Demonstration of Longitudinal Stacking in the ESR with Barrier Buckets and Stochastic Cooling, in \emph{Proceedings of COOL'11}, edited by I. Meshkov (Alushta: JINR, 2011), p. 140

\bibitem{ref7}C. M. Bhat, Applications of barrier bucket RF systems at Fermilab, in \emph{PRIA2006}, edited by K. Takayama (Tsukuba: KEK, 2006), p. 156

\bibitem{ref8}M. Fujieda, Y. Iwashita, A. Noda et al, Phys. Rev. ST Accel. Beams, 2: 122001 (1999)

\bibitem{ref9}http://www1.jinr.ru/Preprints/2013/eng\_053(D9-2013-53).pdf, retrieved 22th March 2016

\bibitem{ref10}H. Poth, Physics Report, 196: 135 (1990)

\bibitem{ref11}A. Hofmann, Tune shifts from self fields and images, in \emph{CAS - CERN Accelerator School: 5th General Accelerator Physics Course}, edited by S. Turner (Jyv$\ddot{a}$skyl$\ddot{a}$: CERN, 1994), p. 329¨C348

\bibitem{ref12}A. V. Smirnov, I. Meshkov, A. O. Sidorin et al, Necessary Condition for Beam Ordering, in \emph{Proceedings of COOL'07}, edited by Volker RW Schaa (Bad Kreuznach: GSI, 2007), p. 87

\bibitem{ref13}A. O. Sidorin, I. N. Meshkov, I. A. Seleznev et al, Nuclear Instruments and Methods in Physics Research Section A, 558: 325 (2006)

\bibitem{ref14}http://betacool.jinr.ru/programs/Betacool\%20Physics\%20Guide.pdf, retrieved 22th March 2016

\bibitem{ref15}V. V. Parkhomchuk, Nuclear Instruments and Methods in Physics Research Section A, 441: 9 (2000)

\bibitem{ref16}L. J. Mao, J. C.Yang, J. W. Xia et al, Nuclear Instruments and Methods in Physics Research Section A, 786: 91 (2015)

\bibitem{ref17}A. Piwinski, Synchro-betatron resonances, in \emph{CAS - CERN Accelerator School: Advanced Accelerator Physics}, edited by S. Turner (Oxford: CERN, 1987), p. 187¨C202

\bibitem{ref18}M. Martini, CERN preprint, CERN-PS-84-9-AA/27

\bibitem{ref19}J.D. Bjorken, S. Mtingwa, Particle Accelerators, 13: 115 (1983)

\bibitem{ref20}S. Abeyratne et al, arXiv:1209.0757




\end{thebibliography}
\end{document}